\documentstyle[sprocl]{article}

\def\al{\alpha}

\def\be{\begin{equation}}
\def\ee{\end{equation}}
\def\bea{\begin{eqnarray}}
\def\eea{\end{eqnarray}}

\def\M{{\cal M}}                       
\def\hs{\qquad}               
\def\nn{\nonumber}            
\def\beq{\begin{eqnarray}}    
\def\eeq{\end{eqnarray}}      
\def\at{\left(}               
\def\aq{\left[}               
\def\ct{\right)}              
\def\cq{\right]}              
\def\R{{\hbox{{\rm I}\kern-.2em\hbox{\rm R}}}}   
\def\H{{\hbox{{\rm I}\kern-.2em\hbox{\rm H}}}}   
\def\N{{\hbox{{\rm I}\kern-.2em\hbox{\rm N}}}}   
\def\C{{\ \hbox{{\rm I}\kern-.6em\hbox{\bf C}}}} 
\def\Z{{\hbox{{\rm Z}\kern-.4em\hbox{\rm Z}}}}   
\def\ii{\infty}                                  
\newcommand{\fr}[2]{\mbox{$\frac{#1}{#2}$}}        
\def\Tr{\mathop{\rm Tr}\nolimits}                  
\renewcommand{\Re}{\mathop{\rm Re}\nolimits}       
\def\dir{/\kern-.7em D\,}                            
\def\lap{\Delta\,}                                 
\def\al{\alpha}
\def\be{\beta}
\def\ga{\gamma}
\def\de{\delta}
\def\ep{\varepsilon}
\def\ze{\zeta}

\def\la{\lambda}

\def\si{\sigma}

\def\Ga{\Gamma}

\begin{document}

\title{ON THE DIMENSIONAL REDUCED THEORIES}

\author{G. COGNOLA}

\address{Dipartimento di Fisica, via Sommarive, 
Povo,\\Trento , Italy\\E-mail: cognola@science.unitn.it} 

\author{S. ZERBINI}

\address{Dipartimento di Fisica, via Sommarive, 
Povo,\\Trento , Italy\\E-mail: ZERBINI@science.unitn.it}

\maketitle\abstracts{
The procedure of the dimensional reduction related to the partition
function of a quantum scalar field living in  curved space-time which is
the warp product of symmetric space is investigated.  
}

In this contribution, we would like to revisited the issue related to 
dimensional reduction, and this is used   when one is dealing 
with quantum fields living on 
 space-times having  some symmetries, namely the 
D-dimensional space-time is the "warp" product 
$M_P \times \Sigma_Q$, where $\Sigma_Q$ is a Q-dimensional 
symmetric space with constant curvature. Such investigation  is mainly 
motivated by recent approaches to black holes physics, initiated in 
\cite{wipf} and continued in \cite{all,all1} and the calculation of the 
effective action after and before the dimensional reduction 
\cite{eli}. 

The idea is very simple: 
since  a generic black hole has a large symmetry in the horizon sector, one 
may consider the two dimensional related reduced theory for which the  
effective action may be obtained functionally integrating the corresponding 
conformal anomaly. This procedure gives rise the problem of the validity of
the approximation and this will be discussed here. A related issue is the 
so called       
dimensional-reduction anomaly \cite{frolov99,sutton}.

Let us consider  a scalar field $\Phi$ propagating in the above 
mentioned space. 
\beq
L_D=-\lap_D+m^2+\xi R_D\,, 
\eeq
in which $m^2$ is 
a possible mass term and $\xi R_D$ a suitable "potential term", 
describing the 
non-minimal coupling with the gravitational field. 
 The "exact" theory, namely the non-dimensional reduced 
one, may be 
described by the path integral (Euclidean partion function)
\beq
Z=\int D\Phi e^{-\int dV_D \Phi L_D \Phi}=e^{-\Ga}\,. 
\eeq
The effective action $\Ga$ has to be regularised and may be  
expressed by means of a zeta-regularised functional determinant 
\cite{ray73-98-154,hawk77-55-133,dowk76-13-3224}
(for recent  reviews, see \cite{eliz94b,byts96-266-1})
\beq
\Ga=-\ln Z=-\frac{1}{2}\aq \ze'(0|L_D)+\ln \mu^2 \ze(0|L_D)\cq\,,
\eeq
$\mu^2$ being the re normalization parameter.
Here, the zeta-function is defined by means 
\beq
\zeta(s|L_D)=\frac{1}{\Ga(s)}\int_0^\ii dt \ t^{s-1}K_t\,,\,\,\,\,
K_t= \Tr e^{-t L_D}\,,
\label{mt}
\eeq
valid for $\Re s> D/2$. Here $ \Tr e^{-t L_D}=\sum_i  e^{-t \la_i}$, $\la_i$ 
being the eigenvalues of $L$. 

One may use other regularisation procedures. 
As an example, the dimensional regularisation is defined by
\beq
\Ga_\ep&=&-\frac{1}{2}\int_0^\ii dt \ t^{\ep-1} \Tr e^{-t L_D}=
-\frac{1}{2}\Ga(\ep)\ze(\ep|L_D)\nn \\
&=&-\frac{1}{2}\at \frac{\ze(0|L_D)}{\ep}+\ze'(0|L_D)+\ga \ze(0|L_D)+
O(\ep)\ct \,.
\label{mtdim}
\eeq
Other regularizations  may be used with $t^{\ep}$ substituted by a 
suitable regularisation function $g_\ep(t)$ (see, for example \cite{kirsten}).
Recall that the zeta-function regularisation is a {\it finite} regularisation
and corresponds to the choice
\beq
g_\ep(t)=\frac{d}{d\ep}\at\frac{t^{\ep}}{\Ga(\ep)}\ct\,.
\eeq
The other ones, as is clear from Eq. (\ref{mtdim}), 
give the same finite part, modulo a re normalization, and contain divergent 
terms as the cutoff parameter $\ep \to 0$ and these divergent terms  have to 
be removed by related counter-terms.

As a consequence, as  will be shown,  a crucial role is played by the 
quantity $\Tr e^{-t L_D}$. With regard to this quantity, its  short-$t$ 
asymptotics  has been extensively studied.
For a second-order operator
on a boundary-less $D$-dimensional (smooth) manifold, it reads
\beq
K_t\simeq \sum_{j=0}^\ii A_j(L_D)\
t^{j-D/2}
\:,\label{tas0}
\eeq
in which $A_j(L_D)$ are the Seeley-DeWitt coefficients, which  can be computed
with different techniques \cite{dewi65b,seel67-10-172}. The divergent terms 
appearing in a generic 
regularisation depend on  $A_j(L_D)$.

In the sequel, we also shall deal with local quantities, which can be defined 
by the local zeta-function. With this regard, it is relevant the local short 
$t$ heat-kernel asymptotics, which reads 
\beq
K_t(L_D)(x)=e^{-tL_D}(x)\simeq \frac{1}{(4 \pi )^{D/2}} 
\sum_{j=0}^\ii a_j(x|L_D)\
t^{j-D/2}
\:,\label{hkl}
\eeq
where $a_j(x|L_D) $ are the local Seeley-DeWitt coefficients. The first ones 
are well known and read 
\beq
a_0(x|L_D)&=& 1\,, \hs 
a_1(x|L_D)=\at -\xi R+-m^2+\frac{R}{6}\ct\,.\nn\\
a_2(x|L_D)&=&\frac{1}{2}(a_1(x|L_D)^{2}+\frac{1}{6} \lap_D a_1(x|L_D)+
c_2(x)\,, 
\label{sde010}
\eeq
where
\beq
c_2(x)=\frac{1}{180}\at \lap_D R+ R^{ijkr}R_{ijkr}-R^{ij}R_{ij}  \ct\,.
\eeq

It may be convenient to re-sum partially this asymptotic expansion and one has
\cite{parker}
\beq
e^{-tL_D}(x)\simeq \frac{e^{t a_1(x|L_D)}}{(4 \pi)^{D/2}}
\sum_{j=0}^\ii b_j(x|L_D)\
t^{j-\fr{D}{2}}
\:.\label{hkl11}
\eeq
The advantage of the latter expansion with respect to the previous one, is due to
the fact that now the expansion $b_j$ coefficients depend on the potential only
through its derivatives. One has 
\beq
b_0(x|L_D)&=&1\,, \hs 
b_1(x|L_D)=0\,, \nn \\
 b_2(x|L_D)&=&-\frac{1}{6}\lap_D V+\frac{1}{36}\lap_D R+c_2(x)\,.
\label{sde01}
\eeq

Since the exact expression of the local zeta-function is known only in a 
limited number of cases, one has to make use of some approximation. If the 
first coefficient $a_1(x|L_D)$ is very large and negative and this is 
true if the mass is very large, 
one may obtain an asymptotics expansion of the local 
zeta-function by means of the short $t$ expansion (\ref{hkl}) and the Mellin 
transform, namely \cite{kirsten}
\beq
\ze(s|L_D)(x)& \simeq& \frac{\Ga(s-\fr{D}{2})}{(4\pi )^{\fr{D}{2}}\Ga(s)}
\at -a_1(x|L_D)\ct^{\fr{D}{2}-s}\nn \\
&+&\sum_{j=2}^\ii
\frac{\Ga(s+j-\fr{D}{2})}{(4\pi )^{\fr{D}{2}}\Ga(s)}
\at-a_1(x|L_D) \ct^{\fr{D}{2}-s-j}b_j(x|L_D)\,.
\label{voros}
\eeq 
The latter expansion directly gives also the analytic 
continuation in the whole complex plane.
The global zeta-function can be obtained integrating over the manifold.

Now, let us  introduce the dimensional-reduced theory according to 
\cite{wipf,frolov99}.
We indicate by $\tilde\M^D$ a $D$-dimensional Riemannian manifold  
with metric $\tilde g_{\mu\nu}$ and coordinates $\tilde x^\mu$ 
($\mu,\nu=1,...,D$)
and by $\M^P$ and $\hat\M^Q$ ($Q=D-P$) two sub-manifolds with 
coordinates $x^i$ ($i,j=1,...,P$) and $\hat x^a$ ($a,b=P+1,...,D$) and
metrics $g_{ij}$ and $\hat g_{ab}$ respectively, related 
to $\tilde g_{\mu\nu}$ by the warped product
\beq 
d\tilde s^2=\tilde g_{\mu\nu}d\tilde x^\mu d\tilde x^\nu=
g_{ij}(x)dx^idx^j+e^{-2\si(x)}\hat g_{ab}(\hat x)d\hat x^ad\hat x^b\:.
\label{metric}\eeq
Here,  $\hat\M^Q=\Sigma_Q$ is a constant curvature symmetric space.  

We shall use the notation $\tilde R^\al_{\be\ga\de}$,
$R^i_{jmn}$ and $\hat R^a_{bcd}$ for Riemann tensors in 
$\tilde\M^D$, $\M^P$ and $\hat\M^Q$ respectively, 
and similarly for all other quantities. 
In the Appendix A, one can find the relationship between the geometrical 
quantities related to the sub-manifolds.

We start with a scalar field $\Phi(\tilde x)$ in the Riemannian 
manifold $\tilde\M^D$. The Laplacian-like operator reads
\beq 
L_D\Phi(\tilde x) =\tilde L\Phi(\tilde x)=(-\tilde\lap+\xi\tilde R+m^2)
\Phi(\tilde x)
=(L+e^{2\si}\hat L)\Phi(\tilde x)\:,
\eeq
where
\beq 
&&L=-\lap+Q\si^k\nabla_k+\xi\aq R+2Q\lap\si-Q(Q+1)\si^k\si_k\cq+m^2\:,
\\
&&\hat L=-\hat\lap \:.
\eeq
In order to dimensionally reduce the theory, let us introduce the harmonic
analysis on $\Sigma_Q$ by means of 
\beq
\hat LY_\al(\hat x)=\la_\al Y_\al(\hat x)\:,
\eeq
$\la_\al,Y_\al$ being the eigenvalues and eigenfunctions 
of $\hat L$ on the symmetric space  $\Sigma_Q=\hat\M^Q$.
For any scalar field in $\tilde\M^D$, we can write
\beq 
\Phi(\tilde x)=\sum_\al\phi_\al(x)Y_\al(\hat x)
\eeq
and for the partition function, after integration over $Y_\al$ in 
the classical 
action,
\beq 
 Z^*&=&\int\:d[\bar\phi]e^{-\int\hat\phi\tilde L\hat\phi\:
\sqrt{\tilde g}\:d^Pxd^Q\hat x}=\prod_\al Z_\al\:,
\eeq
where 
\beq 
Z_\al=\int\:d[\bar\phi_\al]e^{-\int\bar\phi_\al L_\al\bar\phi_\al\:d^Px\:}\:.
\eeq
Here $\bar\phi=\sqrt[4]{\tilde g}\phi$ and $\bar\phi_\al=\sqrt[4]{g}\phi_\al$ 
are scalar densities of weight $-1/2$ and the dimensional reduced operators
read
\beq 
&&L_\al=-\lap+V+e^{2\si}\la_\al\:,
\nn\\
&&V=m^2+\xi\aq R+2Q\lap\si-Q(Q+1)\si^k\si_k\cq
-\frac{Q}2\lap\si+\frac{Q^2}{4}\si^k\si_k\:.
\eeq 
In the following, we will denote by an asterix all the quantities associated 
with 
the dimensional reduced operators. As a result, we formally have
\beq
Z^*=\prod_\al \at \det\frac{L_\al}{\mu^2} \ct^{-1/2}\,.
\eeq

If we {\it ignore} the multiplicative anomaly associated with functional 
determinants, namely the fact that $\ln \det AB \neq \ln \det A+
\ln \det B$ for regularized functional determinants \cite{elizalde}, we have  

\beq
\Ga^*=-\ln Z^*=\frac{1}{2}\sum_\alpha \ln \det\frac{ L_\alpha}{\mu^2}\,.
\label{redac=}
\eeq
This formal expression  may be regularised and renormalized and we have
\beq
\Ga^*_\ep=-\frac{\mu^{2\ep}}{2}\sum_\al \int_0^\ii dt t^{-1} g_\ep(t)
\Tr e^{-tL_\al}\,.
\eeq
Removing the cutoff and, for example  making use of a finite regularisation,
one arrives at
\beq
\Ga^*=\frac{1}{2}\sum_\alpha \zeta'(0|\frac{ L_\alpha}{\mu^2})\,.
\label{redac==}
\eeq
Within this procedure, a quite natural  definition of the  
dimensional-reduction anomaly is \cite{frolov99} 
\beq
A_{DRA}= \Ga-\Ga^* \,.
\label{redeff}
\eeq

However, there exists {\it another} possible procedure: 
 if we {\it do not remove} the ultraviolet cutoff $\ep$, 
we may interchange 
the harmonic sum and the integral and arrive at
\beq
\Ga^*_\ep=-\frac{\mu^{2\ep}}{2} \int_0^\ii dt t^{-1}g_\ep(t)
 K_t^*\,,
\label{redhk*}
\eeq
where we have introduced the dimensionally reduced heat-kernel trace  
\beq
K_t^*=\sum_\alpha \Tr e^{-t L_\alpha}\,. 
\label{redheat}
\eeq

It is clear that within  this second procedure, the existence of a non 
vanishing 
dimensional 
reduction anomaly is strictly related to the fact whether the identity
\beq 
K_t^*=K_t
\label{=kt}
\eeq
holds. 
In the following the validity of the identity
(\ref{=kt}) will be discussed.

First, if the whole space-time (its Euclidean version) is a symmetric space, 
it is quite easy to show that Eq. (\ref{=kt}) is true. The reason is that 
in this case, one has at disposal besides the dimensional reduced one, the 
 total harmonic sum  (see,  for example \cite{camporesi,byts96-266-1}).   

 In general, we shall restrict ourselves to the class of 
non-trivial 
warped space-time already considered and make use of the 
short $t$ heat-kernel expansion. For the exact theory we have (here 
$L_D=\tilde L$)   
\beq 
K_t(\tilde L)=\Tr e^{-t \tilde{L}}&\sim& \frac1{(4\pi t)^{D/2}}
\sum_{n=0}^\ii\:\tilde a_n(\tilde x|\tilde L)t^n\:,
\eeq
with
\beq 
\tilde a_1&=&a_1+e^{2\si}\hat a_1
-\frac{Q}6\aq\lap\si-\at\frac{Q}2-1\ct\si^k\si_k\cq\:,
\label{ta1}\\
\tilde a_2&=&a_2+e^{4\si}\hat a_2+e^{2\si}a_1\hat a_1
-\frac1{90}\si^k\nabla_k R-\frac1{45}...
\label{ta2}\eeq
where all quantities with tilde refers to the whole
manifold $\tilde\M^D$ and all quantities with hat refers to the
sub-manifold $\hat\M^Q$.  

With regard to the dimensional reduced kernel 
\beq 
K_t^*(\tilde L)&=&\sum_\al\Tr e^{-tL_\al}\:,\label{KtLal}
\eeq
where
\beq 
&&L_\al=-\lap+V+e^{2\si}\la_\al\:,
\nn\\
&&V=m^2+\xi\aq R+2Q\lap\si-Q(Q+1)\si^k\si_k\cq
-\frac{Q}2\lap\si+\frac{Q^2}{4}\si^k\si_k\:,
\eeq 
the short $t$ expansion can be computed by means of a  
straightforward (but tedious) computation \cite{cognola2000}, and the result is
\beq 
K_t^*(\tilde L)&\sim&\frac{1}{(4\pi t)^{D/2}}
\sum_{n=0}^\ii\:\tilde a^*_n(\tilde x|\tilde L)t^n
\:.
\label{KtExp}\eeq
where
\beq 
 a^*_1(\tilde x|\tilde L) = \tilde a_1\:,
\label{c1}\eeq
\beq
a^*_2(\tilde x|\tilde L)=\tilde a_2\:.
\label{c2}\eeq

As a consequence, one has

\beq
K_t(\tilde{L})\simeq \frac{e^{t \tilde{a}_1(\tilde x)}}{(4 \pi t)^{D/2}}
\aq 1+ b_2(\tilde x|\tilde L)t^2+ b_3(\tilde x|\tilde L)t^3+... \cq
\:,\label{hklp}
\eeq
\beq
K_t^*(\tilde{L})\simeq \frac{e^{t \tilde{a}_1(\tilde x|L_D)}}{(4 \pi t)^{D/2}}
\aq 1+ b_2(\tilde x|\tilde L)t^2+ b_3^*(\tilde x|\tilde L)t^3+... \cq
\:.\label{hklp*}
\eeq

Thus, it is quite natural to  make the  conjecture that  
$b_n^*(\tilde x|\tilde L)=b_n(\tilde x|\tilde L)$ for every $n$ and Eq. 
(\ref{=kt}) holds exactly. Let us discuss about the 
consequences of this fact. 

After the dimensional reduction, as far as the
effective action is concerned, the 
operation of renormalization
 (addition of couterterms and remotion of the cutoff) and the evaluation of the
 harmonic sum 
{\it do not} commute. If we keep fixed and non vanishing   the 
regularisation parameter, we may perform the harmonic sum, and if 
(\ref{=kt}) holds,  
we  may reconstruct the exact partition function, after renormalization. 
In such case, it is evident that no dimensional reduction 
anomaly occurs. 

On the other hand, one may remove the cutoff, adding the necessary 
couterterms  or using a finite regularisation like the zeta-function one
and perform the harmonic sum at the end. 
In this case, as stressed in reference \cite{frolov99}, one 
has to correct the result by adding dimensional 
reduction anomaly terms.
The reason of this possible discrepancy has been  explained in \cite{frolov99}
as  mainly due to the necessity of the regularisation and renormalization 
of the effective action in spaces with different dimensions. There, it has 
also been   observed a possible connection with the multiplicative 
anomaly. 

Regarding this issue, there exists also a mathematical reason for the 
necessity of these reduction 
anomaly terms. In fact, the harmonic sum of the renormalized dimensionally 
reduced effective action diverges and the dimensional anomaly reduction terms
are also  necessary to recover the exact and {\it finite} result.     
This fact stems  also from the necessity of  the presence of the 
multiplicative 
anomaly, since it also diverges, 
being associated with a product of an  
infinite number of dimensional reduced operators.  

It may be convenient to illustrate the dimensional reduction procedure in the
simplest example one can deal with, namely a free massive scalar field in the 
Euclidean version of the D-dimensional Minkoswki space-time.
We may decompose $R^D=R \times R^{D-1}$, thus $\hat{M}=R^{D-1}$, and $\si=0$,
and $L_{\vec k}=-\partial^2_{\tau}+m^2+(\vec k)^2$. It is easy to show that 
(\ref{=kt}) holds, since 
\beq
\Tr e^{-tL{\vec k}}=V(R^{D-1})e^{-tm^2}\frac{e^{-t\vec k^2}}{(4\pi t)^{1/2}}\,.
\eeq  
For $D$ odd, the exact regularized partition function is
\beq
\Ga=-\frac{V(R^{D})\Ga(-D/2)}{2(4\pi )^{D/2}}m^D\,.
\eeq
On the other hand, the partial reduced effective actions are
\beq
\Ga_{\vec k}=-\frac{V(R)\Ga(-1/2)}{2(4\pi )^{/2}}(m^2+\vec k^2)^{1/2}\,.
\eeq
Thus, $\Ga^*=\sum_{\vec k}\Ga_{\vec k}$ is badly divergent. However, it is 
possible to show that the finite part of this divergent integral reproduces 
$\Ga$. In this particular case, the dimensional reduction anomaly must cancel 
the divergent part. 

As a consequence, any approximation \cite{all} based on the truncation in the 
harmonic 
sum of the dimensional reduced theory, may lead, with regard to the 
comparison with the exact theory,  to incorrect conclusions 
(see also the discussions 
and further references reported in \cite{all1}).


\noindent 







\noindent





\begin{thebibliography}{10}


\bibitem{wipf}
V. Mukhanov, A. Wipf and A. Zelnikov.
Phys. Lett. {\bf B 332}, 283(1994).

\bibitem{all}
R. Bousso and S. Hawking.
Phys. Rev.  {\bf D56}, 7788 (1997);
T. Chiba and M. Siino.
Mod. Phys. Lett.  {\bf A 12}, 709 (1997);
W. Kummer, H. Lieb  and D.V. Vassilevich.
Mod. Phys. Lett.  {\bf A 12}, 2683 (1997);

\bibitem{all1}
S. Nojiri and S. D. Odintsov.
Mod. Phys. Lett. {\bf A 12}, 2083 (1997);
S. Nojiri and S. D. Odintsov.
Phys. Rev. {\bf D57}, 2363 (1998);
S. Nojiri and S. D. Odintsov.
Phys. Rev. {\bf D57}, 4847 (1998);
S.J. Gates, S. Nojiri, T. Kadoyosi and S. D. Odintsov.
Phys. Rev. {\bf D58}, 084026  (1998);
W. Kummer, H. Lieb  and D.V. Vassilevich.
Phys. Rev.  {\bf D58}, 108501 (1998);
S. Nojiri, O. Obregon, S. D. Odintsov and K.E. Osetrin .
Phys. Rev. {\bf D60}, 024008  (1999);
R. Balbinot and A. Fabbri
Phys. Rev.  {\bf D59}, 044031 (1999);
R. Balbinot and A. Fabbri
Phys. Lett.  {\bf B459}, 112 (1999);
R. Balbinot, A. Fabbri and I. Shapiro.
Phys. Rev. Lett.   {\bf 83}, 1494 (1999);
R. Balbinot, A. Fabbri and I. Shapiro.
Nucl. Phys.   {\bf B559}, 301 (1999);
W. Kummer, H. Lieb  and D.V. Vassilevich.
Phys. Rev.  {\bf D60}, 084021 (1999);
F. C. Lombardo, F. D. Mazzitelli  and J. G. Russo.
Phys. Rev. {\bf D59}, 064007 (1999);
 
\bibitem{eli}
E. Elizalde, S.  Naftulin and S. D. Odintsov.
Phys. Rev. {\bf D49}, 2852 (1994);
S. Nojiri and S. D. Odintsov.
Phys. Lett {\bf B463}, 57 (1999).
F

\bibitem{frolov99}
V. Frolov, P. Sutton and  A. Zelnikov. 
Phys. Rev. { D 61}, 02421, (2000).

\bibitem{sutton}
P. Sutton. 
Phys. Rev. { D 62}, 044033, (2000).


\bibitem{ray73-98-154}
D.B.~Ray and I.M.~Singer.
Ann. Math. {\bf 98}, 154 (1973).

\bibitem{hawk77-55-133}
S. W. Hawking.
Commun. Math. Phys. {\bf 55}, 133 (1977).

\bibitem{dowk76-13-3224}
J.S.~Dowker and R.~Critchley.
 Phys.~Rev. {\bf D 13}, 3224 (1976).

\bibitem{eliz94b}
E.~Elizalde, S.~D.~Odintsov, A.~Romeo, A.A.~Bytsenko and S.~Zerbini.
{\em Zeta Regularization Techniques with Applications}.
World Scientific, Singapore (1994).

\bibitem{byts96-266-1}
A.A.~Bytsenko, G.~Cognola, L.~Vanzo and S.~Zerbini.
 Phys.~Rep. {\bf 266}, 1 (1996).



\bibitem{kirsten}
 G.~Cognola, K. Kirsten and S. Zerbini.
 Phys. Rev.  {\bf D 48}, 790 (1993).


\bibitem{dewi65b}
B.S.~DeWitt.
{\em The Dynamical Theory of Groups and Fields}.
Gordon and Breach, New York (1965).

\bibitem{seel67-10-172}
R.T.~Seeley.
 Am.~Math.~Soc.~Prog.~Pure Math. {\bf 10}, 172 (1967).


\bibitem{parker}
L. Parker   and D.J. Toms.
 Phys. Rev.  {\bf D 31}, 953 (1985).



\bibitem{elizalde}
E.~Elizalde, L. Vanzo and S. Zerbini.
 Commun.~Math.~Phys. {\bf 194}, 613 (1998).


\bibitem{camporesi}
R. Camporesi
 Phys.~Rep. {\bf 196}, 1 (1990).




\bibitem{cognola2000}
G. Cognola and S. Zerbini.
{\em On the dimensional reduction procedure}.
hep-th/0008061(2000). 



\end{thebibliography}
\end{document}